\documentclass[thmsa,12pt]{article}
\usepackage{amssymb}

\usepackage{amsmath}
\usepackage{sw20lart}

\input{tcilatex}
\input tcilatex

\begin{document}

\title{Experimental realization of hyper-entangled\\
two-photon states}
\author{C. Cinelli, M. Barbieri, F. De Martini and P. Mataloni \\
Dipartimento di Fisica and \\
Istituto Nazionale per la Fisica della Materia,\\
Universit\`{a} di Roma \textit{La Sapienza},\\
Roma, 00185 - Italy\\
e-mail: paolo.mataloni@uniroma1.it}
\date{}
\maketitle

\begin{abstract}
We report on the the experimental realization of hyper-entangled two photon
states, entangled in polarization and momentum. These states are produced by
a high brilliance parametric source of entangled photon pairs with peculiar
characteristics of flexibility in terms of state generation. The quality of
the entanglement in the two degrees of freedom has been tested by multimode
Hong-Ou-Mandel interferometry.
\end{abstract}

\section{Introduction}

The modern science of quantum information (QI) squarely lies on the concept
and on the applications of entanglement which represents the basic process
underlying quantum computation \cite{1}, quantum teleportation \cite{2} and
some quantum cryptographic schemes \cite{3}. The most successful and
reliable applications of entanglement, such as, for instance, quantum dense
coding \cite{4}, teleportation \cite{5}, entanglement swapping \cite{6},
have been obtained so far in quantum optics, by exploiting the process of
Spontaneous Parametric Down Conversion\ (SPDC) \cite{7}. Here a pair of
photons, at wavelengths (wl) $\lambda _{1}$ and $\lambda _{2}$ and momenta $%
\hslash \mathbf{k}_{1}$ and $\hslash \mathbf{k}_{2}$, may be generated by a
nonlinear (NL)\ optical crystal shined by a pump laser beam at wl $\lambda
_{p}$. The conservation of energy, $\lambda _{1}^{-1}+\lambda
_{2}^{-1}=\lambda _{p}^{-1}$, and of the momentum, $\mathbf{k}_{1}+\mathbf{k}%
_{2}=\mathbf{k}_{p}$, leads to frequency and $\mathbf{k}-$vector correlation
of the emitted photons. The electromagnetic modes are associated with qubits
which are generally encoded by the field polarization. In this framework the
bi-partite $2\times 2$ Hilbert space is spanned by the four maximally
entangled Bell-state basis, 
\begin{equation}
|\Psi ^{\pm }\rangle =\frac{1}{\sqrt{2}}\left( |H_{1},V_{2}\rangle \pm
|V_{1},H_{2}\rangle \right) \text{,}  \label{1a}
\end{equation}
\begin{equation}
|\Phi ^{\pm }\rangle =\frac{1}{\sqrt{2}}\left( |H_{1},H_{2}\rangle \pm
|V_{1},V_{2}\rangle \right) \text{,}  \label{1b}
\end{equation}
expressed in the horizontal ($H$) and vertical ($V$)\ polarization basis.

It is worth noting that the above referred QI schemes \cite{2,4,5,6} can be
completely realized only if the entire set of Bell state measurements is
performed by distinguishing among the four orthogonal Bell states with 100\%
efficiency. Unfortunately enough, this is impossible with standard linear
optics, and the maximum attainable value of efficiency is 50\% \cite{8}.

Entangled states, prepared in more than one degree of freedom, or
hyper-entangled states, have been proposed for optical Bell state analysis,
in particular by entangling two photons in polarization and momentum \cite{9}%
. Different schemes have been proposed which are based, as SPDC source,
either on type II non collinear phase matching \cite{10,11}, or on the type
I two crystal system \cite{12,13}. However, the experimental realization of
this kind of hyper-entangled states has never been realized so far.

In the present work we report on the realization of the simultaneous
polarization and momentum entanglement of photon pairs generated by a high
brilliance SPDC source, recently developed by us \cite{14,15}, with peculiar
characteristics of flexibility in terms of state generation.

\section{Experimental set-up}

The high brilliance source of entanglement is sketched in Fig. 1 \cite{14,15}%
. A Type I, $.5mm$ thick, $\beta $-barium-borate (BBO) crystal is excited in
two opposite directions by a back-reflected $V$-polarized mode-locked
Ti-Sapphire femtosecond laser ($P=400mW$, rep. rate $=76MHz$) operating at
the second harmonic ($\lambda _{p}=397.5nm$) with wavevector $-\mathbf{k}%
_{p} $, i.e. directed towards the left in Fig. 1. The emitted radiation ($%
\lambda =795nm$) and the laser beam are then back-reflected by a spherical
mirror $M$ with radius $R=15cm$, highly reflecting$\ $both $\lambda $ and $%
\lambda _{p}$, placed at a distance $d=R$\ from the NL crystal. A zero-order 
$\lambda /4$ waveplate (wp), placed between $M\ $and the BBO, performs the $%
|HH\rangle \rightarrow |VV\rangle $ transformation on the $2-$photon state
belonging to the left-cone while leaving in its original polarization state
the pump beam ($\lambda _{p}=\lambda /2$) \cite{15}. The back-reflected
laser beam excites an identical albeit distinct downconversion process with
emission of a new radiation cone directed towards the right in Fig. 1 with
axis $\mathbf{k}_{p} $. In this way, the state of the overall radiation,
resulting from the two overlapping indistinguishable cones, is then
expressed by the pure entangled state: 
\begin{equation}
|\Phi \rangle =\tfrac{1}{\sqrt{2}}\left( |H_{1},H_{2}\rangle +e^{i\theta
}|V_{1},V_{2}\rangle \right)  \label{phi}
\end{equation}
In the present source the particular configuration of single-arm
interferometer allows to overcome many of the instability problems due to
the typical phase fluctuations of a standard two arm interferometer. Phase
stability arises from the fact that the superposition of SPDC emission cones
takes place with phase $(0\leq \phi \leq \pi )$\ reliably controlled by
micrometric displacement $(0\leq \Delta d\leq 70\mu m)\ $of the spherical
mirror $M$ along $\mathbf{k}_{p}$ \cite{15}.

A positive lens transforms the overall conical\textit{\ }emission
distribution into a cylindrical\textit{\ }one whose transverse circular
section identifies the so-called entanglement-ring (e-ring), with diameter $%
D=1.6cm$.

The insertion of a zero-order $\lambda /2$ wp in the signal or the idler arm
(cfr. Fig. 2) makes possible to locally transform Eq.(\ref{phi}) into the
state:

\begin{equation}
|\Psi \rangle =\tfrac{1}{\sqrt{2}}\left( |H_{1},V_{2}\rangle +e^{i\theta
}|V_{1},H_{2}\rangle \right) \text{.}  \label{psi}
\end{equation}
Then, by setting $\theta =0\ $or $\theta =\pi $ the state can be locally
transformed in any one of the four Bell states, Eqq. (\ref{1a}) and (\ref{1b}%
).

By this scheme it is possible to manipulate the superposition state by
acting separately on one of the two emission cones. In particular, by a
simple \textit{patchwork} technique, pure states can be easily transformed
into mixed states with various degree of mixedeness. Relevant classes of
quantum states as the Werner States and the Maximally Entangled Mixed States
(MEMS) have been created by this source \cite{16,17}.

Besides polarization entanglement, in the present experiment, momentum
(path) entanglement of two photons has been performed by selecting two pairs
of symmetric points of the e-ring, $a_{1}-b_{2}$ and $a_{2}-b_{1}$ (Fig. 1:
inset), following the idea proposed in Ref. \cite{13}. The diameters
connecting $a_{1}$to $b_{2}$ and $a_{2}$ to $b_{1}$ intercross at the angle $%
\alpha $ (see Fig. 1). The unitary character of the SPDC transformation
allows the biphoton state to keep the same phase of the pump beam,
regardless the value of $\alpha $. Hence the following expression of the
momentum entangled Bell states holds for either one of the two emission
cones: 
\begin{equation}
|\psi ^{\pm }\rangle =\tfrac{1}{\sqrt{2}}\left( |a_{1},b_{2}\rangle \pm
|b_{1},a_{2}\rangle \right) \text{.}  \label{mom}
\end{equation}
In this experiment the phase is setted, $\phi =0$, $\phi =\pi $, by the
insertion of two thin glass plates intercepting the pairs $a_{1}-b_{2}$ and $%
a_{2}-b_{1}$, and by a suitable tilting of one of the two plates (Fig. 1:
inset). Several masks with different values of $\alpha $, each one with four
holes of diameter $d=1.5mm$, corresponding to the points $a_{1}$, $a_{2}$, $%
b_{1}$, $b_{2}$, have been tested..They are mounted on a rotation stage to
optimize the quality of the momentum entanglement.

The SPDC outcoming radiation is divided along a vertical axis in two sets of
modes, $a_{1}-b_{1}$ and $a_{2}-b_{2}$, by a prism-like two-mirror system
and then reflected towards a multimode Hong-Ou-Mandel (HOM) interferometer 
\cite{18} (Fig. 2). Such a measurement appears to be the most direct one to
test the entanglement character of the state in the two degrees of freedom
of polarization and momentum. In the inset of Fig. 2 it is shown how the
modes $a_{1}$ and $b_{1}$ recombine with the corresponding modes $a_{2}$ and 
$b_{2}$ on the plane of a nonpolarizing $50-50$\ beam splitter (BS). A
trombone mirror assembly mounted on a step by step translation stage allows
the fine adjustment of the path delay $\Delta x$ between the mode sets $%
a_{1}-b_{1}$ and $a_{2}-b_{2}$.

The outcoming signals belonging to the output BS modes, $a_{1}^{\prime
}-b_{1}^{\prime }$ and $a_{2}^{\prime }-b_{2}^{\prime }$, are focused on the
active surfaces of two independent avalanche single photon detectors, mod.
SPCM-AQR14. Two equal interference filters, with bandwidth $\Delta \lambda
=3nm$, placed in front of the detectors, determine the coherence-time of the
emitted photons:\ $\tau _{coh}$\textit{\ }$\approx 400f\sec .$

\section{Experimental results:}

\subsection{a) Polarization entanglement}

We have characterized first the polarization entangled states $|\Psi ^{\pm
}\rangle $ generated by the source. For this measurements the four hole mask
has been rotated in order to get the SPDC radiation passing only through the
symmetric holes $a_{1}$and $b_{2}$ (or, equivalently, $a_{2}$and $b_{1}$)
alligned in the $H$\ direction .When operating in the femtosecond regime the
source is affected by a large temporal walk-off effect between the $H$\ and $%
V$ polarization components. It comes out that, because of double passage
through BBO and the $\lambda /4$\ wp, the product state $|V_{1},V_{2}\rangle 
$, corresponding to the SPDC radiation generated toward the left of Fig. 1,
is advanced by $\sim 540f\sec $ with respect the state $|H_{1},H_{2}\rangle $%
, generated toward the right. As a consequence, due to the temporal width of
the biphoton wavepacket, the two polarization components do not overlap.
Temporal indistinguishability is recovered by means of a $18mm$ quartz plate
($Q$) intercepting the entire e-ring, aligned at the output of the source
with the optic axis oriented along the $H$ direction (Fig. 2).

We have tested the polarization entanglement by means of two standard
polarization analyzer settings (not shown in Fig. 2) in front of the
detectors. Without the BS in Fig. 2, a polarization interference visibility
of $\sim 90\%$ has been measured for the Bell state $|\Phi ^{-}\rangle $ 
\cite{19}.

By inserting the BS and removing the polarization analyzers, standard HOM
interferometric tests have been performed for the Bell states $|\Psi ^{\pm
}\rangle =\frac{1}{\sqrt{2}}\left( |H_{1},V_{2}\rangle \pm
|V_{1},H_{2}\rangle \right) $, corresponding to two different positions of
mirror $M$ with respect the NL crystal. The experimental results given in
Fig. 3a, clearly indicate the typical dip-peak interference due to the
entangled nature of the state, corresponding to a visibility: $V=(0.87\pm
0.01)$. The FWHM $(\simeq 60\mu m)$ of the interference pattern is in
agreement with the expected value for a filter bandwidth $\Delta \lambda
=3nm $. We have also tested the coincidence fringe visibility for a path
length difference $\Delta x=0$, by varying the phase $\theta $ with the
position of mirror $M$. The results given in Fig. 3b show the expected
interference behaviour of the number of coincidences $N_{C}(\left| \Psi
\right\rangle )\varpropto 1-\cos \theta $, with periodicity $\sim 70\mu m$ 
\cite{20}.

\subsection{b) Momentum entanglement}

The momentum entangled states $|\psi ^{\pm }\rangle $ have been investigated
with the four hole mask adjusted as shown in Fig. 1. The radiation belonging
to either one of the emission cones of the source (cfr.Fig. 1) has been
detected and measured. Different values of the angle $\alpha $, ranging from 
$10%
{{}^\circ}%
$ to $40%
{{}^\circ}%
$, have been tested. In all the cases the typical Hong-Ou-Mandel dip
behaviour has been observed as a function of $\Delta x$ for the state $|\psi
^{+}\rangle $, $\phi =0$ in Eq.(\ref{mom}).

All the possible falsification tests have been performed to prove the
entanglement character of the state. In particular we have found that the
dip disappears when either one of the mode pairs, $a_{1}-b_{2}$, or $%
a_{2}-b_{1}$ is covered. On the other hand no coincidence is observed when
either the hole pairs of the rotating mask $a_{1}-a_{2}$ or $b_{1}-b_{2}$
are covered.

The main proof of the existance of momentum entanglement is given by the
possibility of manipulating the phase of the state $|\psi \rangle $, viz. by
tilting one of the glass plates of Fig. 1. Similarly to the case of
polarization entanglement, Fig. 4a shows the triplet-singlet transition, $%
\phi =0\rightharpoonup \phi =\pi $, obtained in the HOM interferometer with
an interference visibility is $V=(0.82\pm 0.01)$. The oscillation behaviour $%
N_{C}(\left| \psi \right\rangle )\varpropto 1-\cos \phi $, obtained for $%
\Delta x=0$, as a function of $\phi $ is given in Fig. 4b \cite{20}.

\subsection{c) Hyper-entanglement}

The hyper-entangled state realized in the present experiment is given by the
factorization of the Bell states representing the entanglement in
polarization and momentum, $\left| \Xi \right\rangle =\left| \Psi ^{\pm
}\right\rangle \otimes \left| \psi ^{\pm }\right\rangle $. It can be
optimized by the simultaneous, independent manipulation of the phases $%
\theta $ and $\phi $. We can express the state $\left| \Xi \right\rangle $
as 
\begin{equation}
\left| \Xi \right\rangle =\frac{1}{2}\left\{ \left| H_{1},V_{2}\right\rangle
+e^{i\theta }\left| V_{1},H_{2}\right\rangle \right\} \left\{
|a_{1},b_{2}\rangle +e^{i\phi }|b_{1},a_{2}\rangle \right\}  \notag
\end{equation}
In terms of the creation operators of the electromagnetic field applied to
the vacuum state, we can write 
\begin{equation}
\left| \Xi \right\rangle =\frac{1}{2}\left\{ a_{1H}^{\dagger
}b_{2V}^{\dagger }+e^{i\theta }a_{1V}^{\dagger }b_{2H}^{\dagger }+e^{i\phi
}b_{1H}^{\dagger }a_{2V}^{\dagger }+e^{i(\theta +\phi )}b_{1V}^{\dagger
}a_{2H}^{\dagger }\right\} \left| 0\right\rangle \text{,}
\end{equation}
where $a_{j\sigma }^{\dagger }$ and $b_{j\sigma }^{\dagger }$ represent the
operators associated to the modes $a_{j}$ and $b_{j}$, respectively, and to
the polarization $\sigma $ ($j=1,2$ and $\sigma =H,V$). In a HOM
interference experiment the operator relations existing between the input
and output modes of the BS \cite{21} allow to express the state $\left| \Xi
\right\rangle $ as 
\begin{equation}
\left| \Xi \right\rangle =\frac{1}{4}\left\{ 
\begin{array}{c}
\left[ 
\begin{array}{c}
a_{1H}^{\dagger ^{\prime }}b_{2V}^{\dagger ^{\prime }}\left( 1-e^{i(\theta
+\phi )}\right) -ia_{2H}^{\dagger ^{\prime }}b_{2V}^{\dagger ^{\prime
}}\left( 1+e^{i(\theta +\phi )}\right) \\ 
-ia_{1H}^{\dagger ^{\prime }}b_{1V}^{\dagger ^{\prime }}\left( 1+e^{i(\theta
+\phi )}\right) -a_{2H}^{\dagger ^{\prime }}b_{1V}^{\dagger ^{\prime
}}\left( 1-e^{i(\theta +\phi )}\right)
\end{array}
\right] + \\ 
+e^{i\theta }\left[ 
\begin{array}{c}
a_{1H}^{\dagger ^{\prime }}b_{2V}^{\dagger ^{\prime }}\left( 1-e^{i(\phi
-\theta )}\right) -ia_{2V}^{\dagger ^{\prime }}b_{2H}^{\dagger ^{\prime
}}\left( 1+e^{i(\phi -\theta )}\right) \\ 
-ia_{1V}^{\dagger ^{\prime }}b_{1H}^{\dagger ^{\prime }}\left( 1+e^{i(\phi
-\theta )}\right) -a_{2V}^{\dagger ^{\prime }}b_{1H}^{\dagger ^{\prime
}}\left( 1-e^{i(\phi -\theta )}\right)
\end{array}
\right]
\end{array}
\right\} \left| 0\right\rangle  \notag
\end{equation}
Hence the expected number of coincidences at the output of HOM is: 
\begin{equation}
N_{C}(\left| \Xi \right\rangle )=\frac{1}{2}(1-\cos \phi \cos \theta )\text{.%
}  \label{hyper2}
\end{equation}
This behaviour can be easily understood by considering that, when $\theta =0$%
, $\phi =0$ or $\theta =\pi $, $\phi =\pi $, the symmetry of the state $%
\left| \Xi \right\rangle $ is bosonic, hence we expect the photons go
together through the same arm of the BS. On the contrary, for $\theta =0$, $%
\phi =\pi $ or $\theta =\pi $, $\phi =0$, the state $\left| \Xi
\right\rangle $ is characterized by a fermionic symmetry and the photons go
separately into the two output arms of the BS.

The experimental results of Fig. 5 verify the bosonic-fermionic transition
predicted for the state $\left| \Xi \right\rangle $\ by the expression (\ref
{hyper2}). In this case the phase $\phi $ has been varied by tilting one of
the glass plates for two given positions of mirror $M$, corresponding to the
phase values $\theta =0$, $\pi $.

An accurate experimental production of the hyper-entangled state has been
found particularly severe, likely because of the general critical
requirements needed for operating on the simultaneous conditions of
entanglement in the two degrees of freedom of polarization and momentum.
This is confirmed by the value of the oscillation visibility, $V\simeq 60\%$
obtained in this experiment. We do believe that the experimental
difficulties due to the need of a simultaneous superposition of the modes in
two different region of BS and a possible unperfect modal structure of the
pump beam may be also responsible of this problem.

\section{Conclusion}

The experimental realization of two photon states, simultaneously entangled
in polarization and momentum, so-called hyper-entangled states, has been
presented in this paper. These states have been realized by selecting two
pairs of correlated $\mathbf{k-}$vectors within the degenerate emission cone
of a high brilliance source of polarization entangled photons, recently
developed by us. Hyper-entangled states are manipulated by varying
independently the phases corresponding to the two entanglement degrees of
freedom of the states. We have investigated the entanglement character of
these states by multimode HOM interferometry. They can be symmetric or
antisymmetric, depending on the combined values of the two phases. The
additional degree of freedom given by the possibility of varying either one
of the two phases, may represent a useful control parameter in quantum state
engineering and Bell state measurements.

This work was supported by the FET European Network on Quantum Information
and Communication (Contract IST-2000-29681: ATESIT), MIUR
2002-Cofinanziamento and PRA-INFM\ 2002 (CLON).

\begin{description}
\item[Fig. 1 - ]  Generation of hyper-entangled two photon states:
experimental setup. Displacement of mirror $M$\ allows to change the phase $%
\theta $. Inset: four hole mask and glass plate system for phase $\phi $\
adjustment..

\item[Fig. 2 - ]  Multimode HOM interferometer. After walk-off compensation (%
$Q$), the mode sets $a_{1}-b_{1}$ and $a_{2}-b_{2}$, travelling along the
two arms of the interferometer, are recombined onto a nonpolarizing
symmetric BS by translation $\Delta x$. Inset: spatial coupling of the input
modes $a_{1}-b_{1}$, $a_{2}-b_{2}$ on the BS. The BS output modes, $%
a_{1}^{\prime }-b_{1}^{\prime }$, $a_{2}^{\prime }-b_{2}^{\prime }$ are also
shown.

\item[Fig. 3 - ]  a) Coincidence rate depending on the path length
difference $\Delta x$ for the polarization entangled states $|\Psi
^{+}\rangle $ ($\theta =0$) and $|\Psi ^{-}\rangle $ ($\theta =\pi $). b)
Coincidence rate measured as a function of the mirror $M$\ position ($\Delta
x=0$).

\item[Fig. 4 - ]  a) Coincidence rate depending on the path length
difference $\Delta x$ for the momentum entangled states $|\psi ^{+}\rangle $
($\phi =0$) and $|\psi ^{-}\rangle $ ($\phi =\pi $). b) Coincidence rate
measured as a function of phase $\phi $, expressed in units of $\pi $. ($%
\Delta x=0$).

\item[Fig. 5 - ]  Coincidence rate vs. $\phi $ (in units of $\pi $) for the
hyper-entangled state $\left| \Xi \right\rangle $ ($\Delta x=0)$. Continuous
line: $\theta =0$, dashed line: $\theta =\pi $.
\end{description}

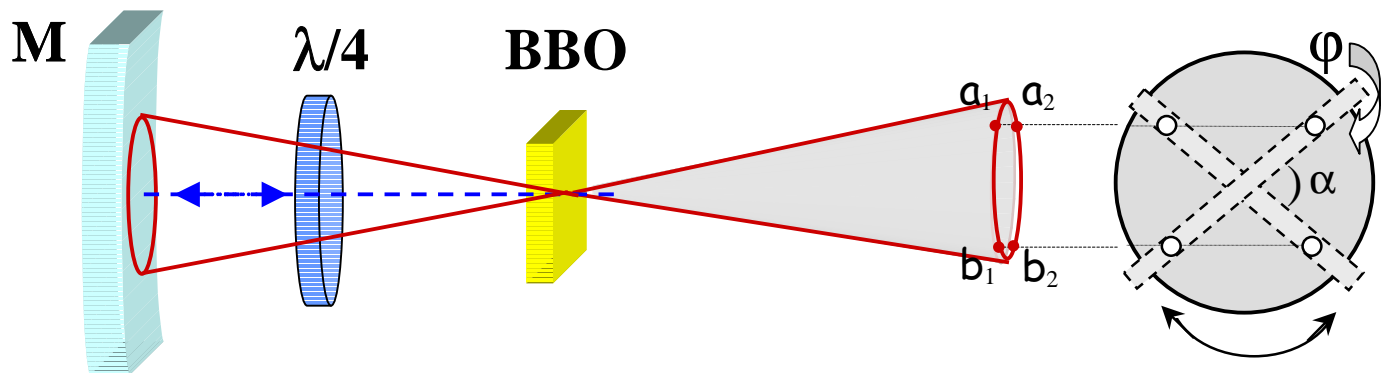
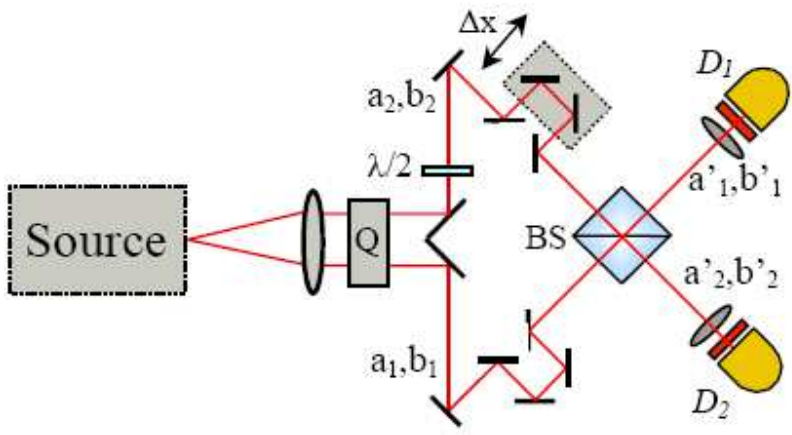
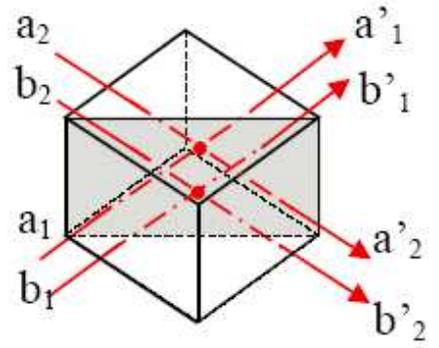
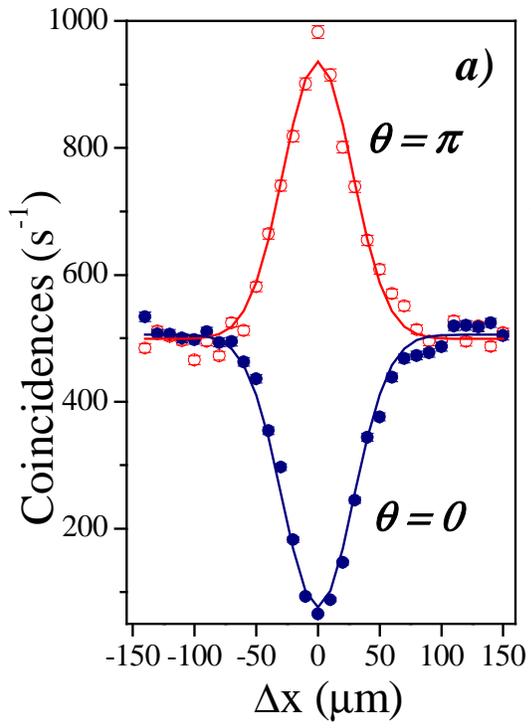
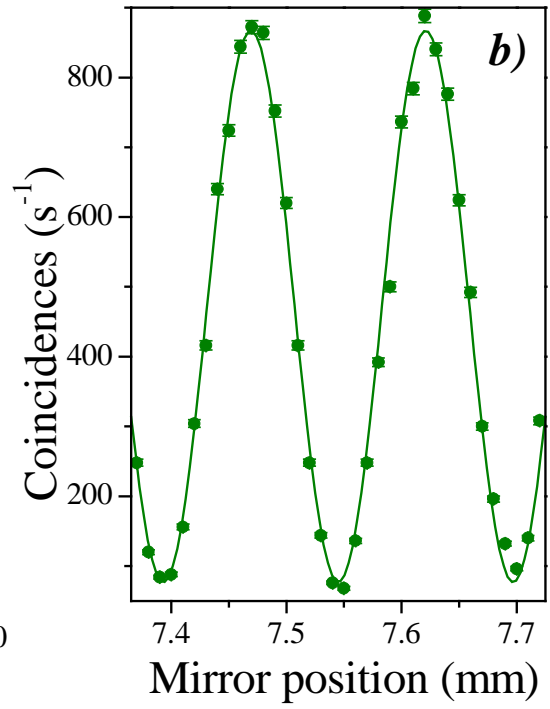
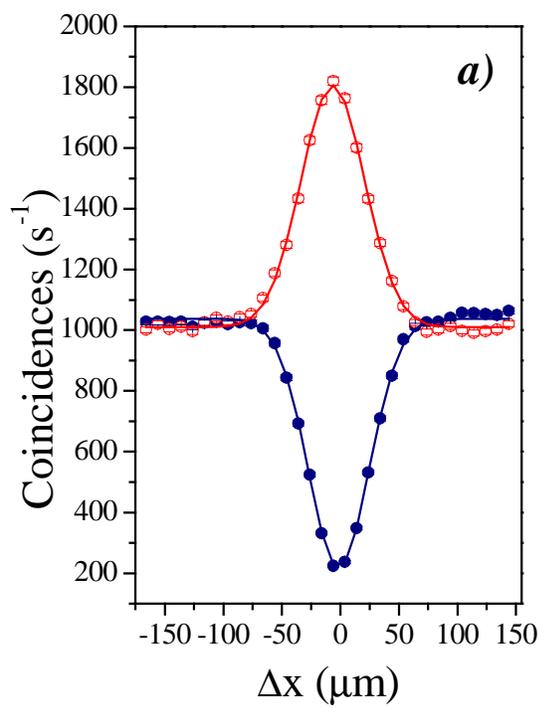
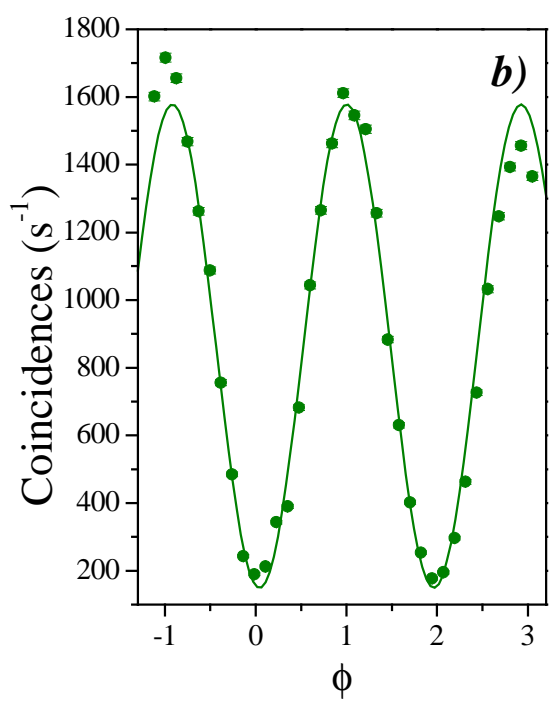
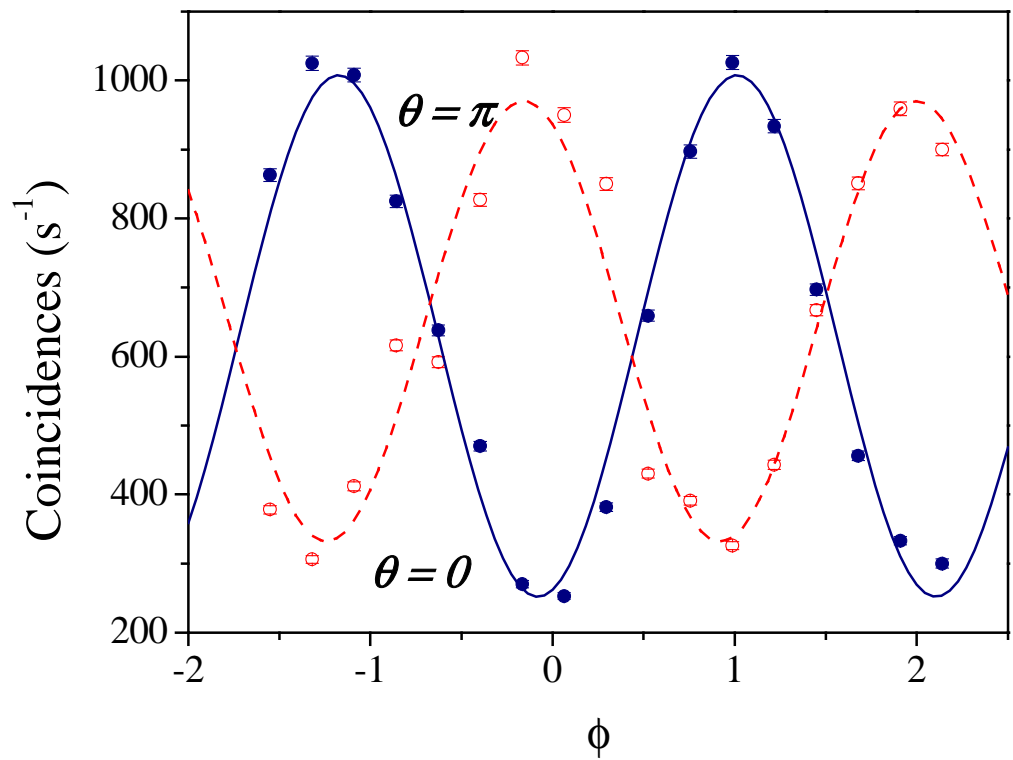
\end{document}